\documentclass[10pt, conference, compsocconf]{IEEEtran}
\ifCLASSINFOpdf
   \usepackage[pdftex]{graphicx}
\else
\fi
\usepackage{algorithmic}
\usepackage{color,soul}
\definecolor{lightblue}{RGB}{240,240,230}
\sethlcolor{lightblue}

\begin{document}
%
\title{Building Internal Cloud at NIC : A Preview}


\author{\IEEEauthorblockN{S.M.M.M Kalyan kumar}
\IEEEauthorblockA{National Informatics Centre\\
 New Delhi, India\\
 Email: kumar.kalyan@nic.in}
\and
\IEEEauthorblockN{Dr S C Pradhan}
\IEEEauthorblockA{National Informatics Centre\\
 New Delhi, India\\
 Email: pradhan@nic.in}

}


%


\maketitle

\begin{abstract}

The most of computing environments in the IT support organization like NIC are designed to run in centralized datacentre. The centralized infrastructure of various development projects are used to deploy their services on it and connecting remotely to that datacentre from all the stations of organization. Currently these servers are mostly underutilized due to the static and conventional approaches used for accessing and utilizing of these resources. The cloud patterns is much needful for optimizing resource utilization and reducing the investments on unnecessary costs. So, we build up and prototyped a private cloud system called $nIC$(NIC Internal Cloud) to leverage the benefits of cloud environment. For this system we adopted the combination of various techniques from open source software community. The user-base of $nIC$ consists developers, web and database admins, service providers and desktop users from various projects in NIC. We can optimize the resource usage by customizing the user based template services on these virtualized infrastructure. It will also increases the flexibility of the managing and maintenance of the operations like archiving, disaster recovery and scaling of resources. The open-source approach is further decreases the enterprise costs. In this paper, we describe the design and analysis of implementing issues on internal cloud environments in NIC and similar organizations.
\end{abstract}


%
\IEEEpeerreviewmaketitle

\section{Introduction}
Cloud computing is a supercomputing model that offers the services which solves the vast kind of user requirements efficiently. Other changes it gives the provisioning to parallel and dynamic processing to the end users and offers virtualized, scalable, on demand resources to the end users over the internet. It eliminates the challenges in non-cloud techniques on scaling up and down of resources, upgrading of hardware and software components and monitoring of services. So, further we discuss the appropriateness and necessity of cloud environments at organizations like NIC [1].
 An Internal cloud aims to deliver many of the characteristics of public cloud computing such as scalability and elasticity, the pooling of shared infrastructure, user self-service, availability and reliability. However, by taking a internal cloud approach, organizations can deliver these goals while still using their private physical resources allowing them to keep up complete control and security over their data and applications. By giving application owners better visibility over their resource usage, organizations are able to more easily apply their strategies to enhance the throughput. A self-service interface to which standardized services are published from IT provider which eases application owners and other internal users are able to easily provision resources dynamically. 
The working of conventional approaches not ideal to centralized patterns would follow the dynamic and non-uniform nature of requirements in resources. The cloud is an effective reuse model where reusable services are deployed once and shared by many potential consumers. So, further we discuss the appropriateness and necessity of cloud environments at organizations like NIC.
 The Internal cloud pattern can enhance the user experience and decreases operational costs with its nature of cloud techniques. It adapt to deal the situations like  sudden increase or decrease of rate of demand of resources and outplay the traditional methods which fails in those situations. We can offer various heterogeneous combinations of software stacks for different [12] project requirements. The project requirements include database, ticket, domain, kerberos, mail, print, middleware, clients, net, storage, build, test, versioning, and so on. The platform can host the different combinations of operating systems and software and provisions the on demand service environment which increases the productivity by offloading the users from these platform configurations. The seamless working state of these platforms from the perspective of users for a medium-sized company like NIC gives provision of considering the solutions for the production environments.
National Informatics Centre(NIC) is a premier Institute and government software agency which is running various software projects at different datacenters spans across the states of India. The datacentres are well-connected by high bandwidth network backbone which will give rich Network services. So, the availability of huge footprint of hardware resources gives the good chance to implementing large-scale cloud environments.
However, implementing the required private cloud architecture at production level datacentres needs the seam-less working software. In the open source community we can find such a software to deploy the private cloud. The infrastructure virtualization components like Xen or XCP [2] and virtual desktop software like XVP [9], storage virtualization/cluster components like SWIFT [5], orchestration components like Cloudstack [3], Openstack [4] gives the wider options to implement the Internal cloud. 
In this paper, we propose internal cloud architecture to implement the infrastructure, Platform and software as a service to the developers, users from various projects [1] of NIC and appropriate maintenance and monitoring techniques to control the system. The rest of the paper is organized as follows. Section 2 describes building of the system. Section 3 describes the design details. Section 4 gives the details deployment and performance. Section 5 concludes the paper.

\section{Building of Internal Cloud}
\subsection{Key Technologies of the $NIC$ Environment}
In this section we present the key technologies of design choice to implement the private cloud model, including virtualization software, user management orchestration software, VDI and storage service technologies. 
\subsubsection{Server Virtualization}
Virtualization layer provides significant benefit for organizations delivering Internal cloud solutions through enhanced scalability and virtual machine mobility. The server virtualization is at the root level of the any cloud setup which segregate or aggregates the computing server pools. The computing pools include central processing units(CPU), memory, disk, and io channels. It works transparently to the hosted application intended to improve stability, utilization, or ease of management of the system. Many variants are available in the virtualization includes hardware virtualization, para virtualization, full virtualization and operating system level virtualization according to the type of resources available. Open source solutions like Xen or XCP gives the a complete enterprise level of virtualization services. It works using the H/W and Para virtualization techniques.  \\
\textbf{Hardware Virtualization} \newline
Hardware supported virtualization is where the cpu has additional hardware support/instructions to facilitate some common tasks usually seen in virtualization. The hardware provides architectural support that facilitates building a virtual machine monitor and allows guest OS's to be run in isolation. \\
\textbf{Para Virtualization} \newline
Paravirtualization is the concept of making changes to the kernel of a guest operating system to make it aware that it is running on virtual, rather than physical, hardware, and so exploit this for greater efficiency or performance or security. It gives the more flexibility and security to the guest instances running in the virtualized platform. \\
\textbf{XCP} \newline
XCP runs as a bare metal hypervisior and consist set of tools to manage the virtual instances running on it. The architecture of the XCP is very simple and gives production ready service to the end users. It can able to create server pools consist of one master nodes and multiple slave nodes. It can scales up and down the servers easily and migrate the vms among the server pool. Rich set of management software also available to configure and monitor the XCP platform like XenCenter, OpenXenManager and XenWebManager.

\subsubsection{Desktop Services}
The centralized virtual servers need to be accessed by the end users in order to use those resources. Desktop virtualization gives the accessing of remote resources and enable IT to deliver the right desktops to meet the needs of every user. With centralized desktop service users data is well protected, provisions rich set of resources and decreases the operational costs. Users can connect to their Virtual instances through the VNC protocol, SSH, or XVP protocol.\\
\textbf{XVP} \newline
The Xen VNC Proxy(XVP) gives the Xen based virtual instance access deployed on XCP machines. XVP gives the proper user management in the private cloud Setup so that users also can access, start and stop the instances in the virtual farms. 
\subsubsection{Storage Services}
The Storage services in cloud are characterized by repeatable, automated provisioning to the end users. Conventional storage devices like SAN, NAS, Local Hard Disks are turned to form the infrastructure for these services and provides consistent, cost-effective solution. Various types of storage services are available like permanent and transient, high and low latency, high and low bandwidth, high and low protected services. Techniques like LVM, Cluster file systems, Gluster FS, DRBD, SWIFT together gives the seamless storage services [13] in cloud environments. \\
\textbf{SWIFT} \newline
Swift is a multi-tenant, highly scalable and durable object storage system that was designed to store large amounts of unstructured data at low cost via a RESTful http API. Swift is used to meet a variety of needs. Swift's usage ranges from small deployments for "just" storing vm images, to mission critical storage clusters for high-volume websites, to mobile application development, custom file-sharing applications, data analytics and private storage infrastructure-as-a-service.  \\
\textbf{DRBD} \newline
DRBD [6] stands for Distributed Replicated Block Device, which replicates data at the block level between two or more sites. DRBD is widely used as high availability and disaster recovery replication technology. DRBD takes over the data, writes it to the local disk and sends it to the other
host. On the other host, it takes it to the disk there. \\
\textbf{GlusterFS} \newline
GlusterFS [7] is a powerful network/cluster filesystem written in user space which uses FUSE to hook itself with VFS layer. GlusterFS takes a layered approach to the file system, where features are added/removed as per the requirement. Though GlusterFS is a file System, it uses already tried and tested disk file systems like ext3, ext4, xfs, etc. to store the data. It can easily scale up to petabytes of  storage which is available to user under a single mount point.\\
\subsubsection{Cloud Orchestration and Networking}
The management of virtual resources is critical step to forming the private cloud. Open source tools like CloudStack, OpenStack gives enough flexibility to deploy the on the fly private cloud in the premises. With use of these tools we provision the self-service portals for virtual resources to the end users and monitoring, metering of usage of resources.  \\
\textbf{CloudStack} \newline
CloudStack can be used for IaaS solution which builds private clouds. It enables compute orchestration, Network-as-a-Service, user and account management, a full and open native API, resource accounting, and a first-class user interface. CloudStack works in monolithic way includes management server take control of total setup. It having two modes of operation as basic for simple network setup and advanced for complicated network setup. \\
\textbf{OpenStack} \newline
Openstack software designed to orchestrate the large networks of virtual machines, gives available, scalable and on-premise cloud infrastructure platform. It consists of various modules to enable the complete IaaS services includes image store provides a catalog and repository for virtual disk images, Compute provides virtual servers upon demand, dashboard provides a modular web-based user interface, identity provides authentication and authorization, Quantum service provides network connectivity as a service, Ceilometer provides metering and block store provides persistent block storage to guest vms. Stack of different technologies used for $nIC$ are presented in fig:1. \\

\begin{figure}[h]
 \centering
 \includegraphics[width=70mm,height=60mm]{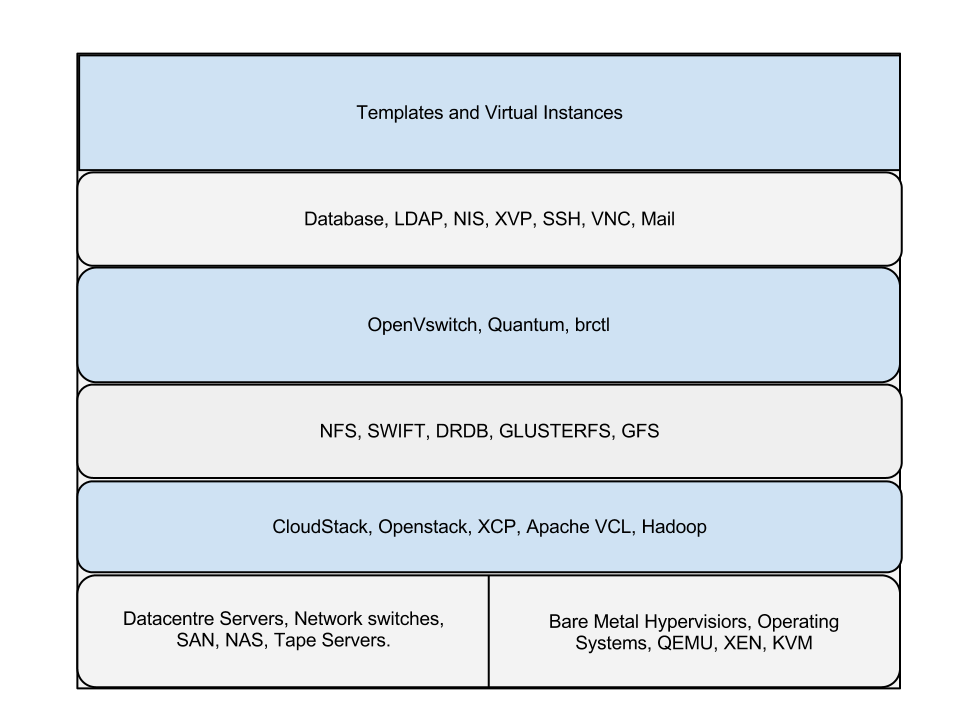}
 \caption{$NIC$ Paradigm }
 \label{fig:1 $nIC$  Paradigm}
\end{figure}

\subsection{Architecture of the $NIC$}

 \subsubsection{Prototype for $NIC$}
       The framework consists of all the components required to establish the total Cloud environment and orchestrates the virtual resources among the user requests. The components are worked at different roles from the client interaction to the actual processing and storing of data. The basic layers for the framework are top layer consists of user interfacing tools like xvpweb, Xencenter, OpenXenManager, CloudstackGUI and Openstack KeyStone. The middle layer consists of hypervisors like Xen, Orchestration software cloudstack and OpenStack, and user management systems like ldap, nis servers. 

The bottom layer consists of stores of virtual machines images, data repositories, storage services like swift, drbd, nfs and glusterfs. All  these layers are loosely coupled and interacted themselves gives the complementary services in the private cloud environment. The private cloud resources can be segregated according to the respective project domains. Upon the project basis we can customize the software stack and given it as a on-the-fly platform for the project development. The layers according to the different functionalities of the system can be seen below: 
     
\textbf{Three layers of the $NIC$} \newline
-------------------------------------------------------------- \newline

    \begin{algorithmic}
      \STATE \textit{\hl{* The Framework/Admin receives the resource requests from cloud clients}}\\
      \STATE \textit{\hl{* Request processed by creating new virtual resource allotments}}\\
      \STATE \textit{\hl{1. Accessing of cloud resources from the top layer of the framework}} \\
      \STATE \textit{\hl{ a. SSH, VNC and RDP protocol access}} \\
      \STATE \textit{\hl{ b. XVP and VDI access}} \\
      \STATE \textit{\hl{2. Computational services from the middle layer of the framework}}\\
      \STATE \textit{\hl{ a. System and Management services}}\\
      \STATE \textit{\hl{ b. Virtual hardware services}}\\
      \STATE \textit{\hl{3. Storage services from the bottom layer of the framework }} \\
      \STATE \textit{\hl{ a. Storage services for on-the-fly computational requests}} \\
      \STATE \textit{\hl{ b. Storage services for archiving requests}} \\
      \STATE \textit{\hl{* Monitoring and Maintenance of production services}}\\
    \end{algorithmic}

----------------------------------------------------------- \newline

\textbf{Project based resource allotment:}\\
        We can divide the server pools as per datacentre based or project based. With in a datacentre there exists multiple projects and allocated physical servers. According the project allotments of the servers we create a customized private cloud on those servers. The shared requirements exists on the multiple projects can be serviced from the single datacentre servers. The disaster recovery site can be placed at geographically different datacentre in order to protect the cloud. 
\begin{figure} [http]
 \centering
 \includegraphics[width=90mm,height=40mm]{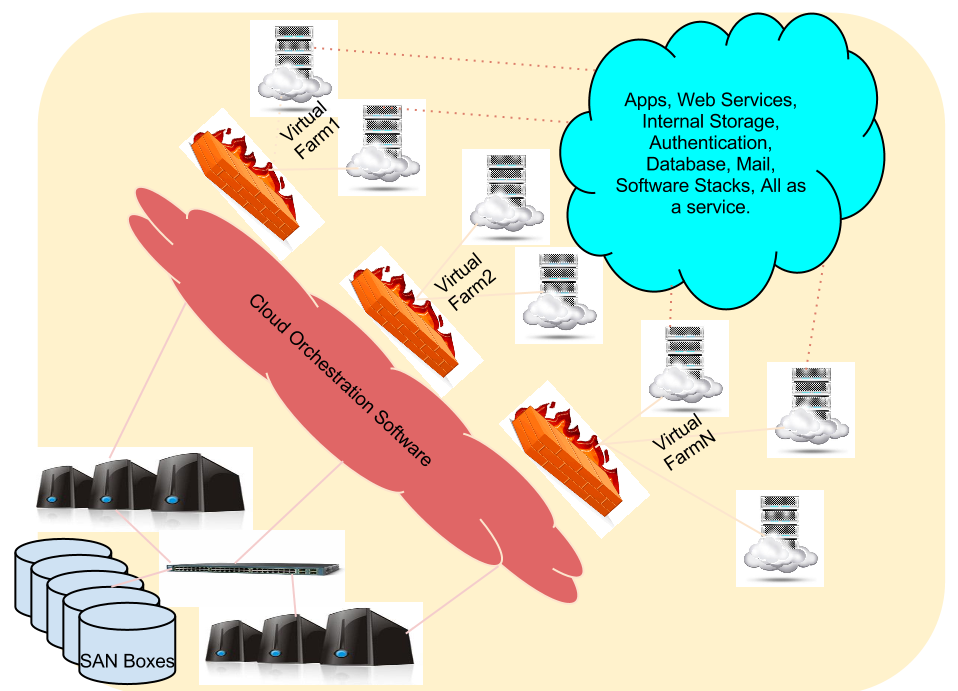}
  \caption{Project based Resource Farms}
 \label{fig:2}
\end{figure}

At the higher level the users see only the dedicated allocation of servers and no have idea of physical placement of virtual servers. The virtual farm is totally work in transperent way to dedicate the reources to end user workloads or projects.

\subsubsection{Template Management}

The users of $nIC$ are from the different projects with different requirements. According to those requirements $nIC$ have the ready made templates to build the on-the-fly resource. The users can build their own template and uses it for own service. The templates are customized with various software stacks needed for project development, deployments etc. The users can instantiate many number of virtual instances from the single template. The life cycle from template creation to virtual instance explained in the following routine. \\

\textbf{Life Process of a Virtual Instance}\newline
----------------------------------------------------------- \newline
\begin{algorithmic}
       \STATE  \textit{\hl{1. Booking of the template from the template store}}\\
       \STATE  \textit{\hl{a. Pre-configured template available in the store}}\\
       \STATE  \textit{\hl{b. On-the-fly Creation of template by the user}}\\
       \STATE  \textit{\hl{2. Attributing the instance}}\\
       \STATE  \textit{\hl{a. Network service assignment}}\\ 
       \STATE  \textit{\hl{b. Storage service assignment}} \\
       \STATE  \textit{\hl{3. Start the instance on server pool}} \\
       \STATE  \textit{\hl{4. Remote service assignment}}\\
       \STATE  \textit{\hl{5. Monitoring and backup service assignment}}\\
       \STATE  \textit{\hl{6. Stop and destroy the instance}}\\
\end{algorithmic}
----------------------------------------------------------- \newline
\subsection{Design of $NIC$}
The design of $nIC$ consists of the all modules specification and its services involved in the internal cloud. The $nIC$ offers the all the dimensions that a cloud consists are infrastructure service, platform service and software service. The module service specification is presented here in the following subsections.
\subsubsection{User Management} 
  There are several people from different projects in the organization that deal with data center and cloud. It is required to define role and assign people to those role and based on their role, they should have different access. Administrators should be able to view and change everything and users should only have access to view everything without modify them. Each project have different requirements of role and access. $nIC$ provides user management capability at different levels of the framework.\\ 

\textbf{Authentication services in $nIC$}\newline
----------------------------------------------------------- \newline
\begin{algorithmic}
       \STATE  \textit{\hl{1. Authenticate with the $nIC$ framework}}\\
       \STATE  \textit{\hl{2. Authenticate with the virtual image service}}\\
       \STATE  \textit{\hl{3. Authenticate with the storage service}}\\
       \STATE  \textit{\hl{4. Authenticate with the network and remote access service}}\\
\end{algorithmic}
----------------------------------------------------------- \newline

\subsubsection{Network management}
The Network services in private cloud offers the isolation, flexibility and self-service among the virtual resources allocated to the end users. The virtual networks assigned to the instances defines the reachability and accessibility of the user service.  The software switches gives the flexibility to tune and configure the performance control on the virtual networks. The software defined firewalls, load balancers gives the flexible tuning of security and performance controls. Dhcp based ip assignment and datacentre level vlans gives rich network service to the cloud users.  \\
\textbf{Virtual Network services in $NIC$}\newline
----------------------------------------------------------- \newline
\begin{algorithmic}
       \STATE  \textit{\hl{1. Assign the virtual networks from network pools}}\\
       \STATE  \textit{\hl{2. Assign the ip and mac address from address pool}}\\
       \STATE  \textit{\hl{3. Assign the firewall rules}}\\
       \STATE  \textit{\hl{4. Assign the network load balancer rules}}\\
\end{algorithmic}
----------------------------------------------------------- \newline
\subsubsection{Storage Management}
Private cloud storage is elastic, automated and multi-tenant. According to service the storage is transient and low latency or permanent and high latency. The virtual instances are created on minimal storage required for the instance to run. For archiving of data object based storage services are used which gives reliable and rapid provisioning of storage services.  \\
\textbf{Virtual storage services in $NIC$}\newline
----------------------------------------------------------- \newline
\begin{algorithmic}
       \STATE  \textit{\hl{1. Assign the block storage}}\\
       \STATE  \textit{\hl{2. Assign the object storage}}\\
       \STATE  \textit{\hl{3. Assign the centralized storage services}}\\
       \STATE  \textit{\hl{4. Assign the cluster and high available storage services}}\\
\end{algorithmic}
----------------------------------------------------------- \newline

\subsubsection{Remote Access Control}
The remote access hosted cloud scenario provides a secure way for users to access resources in the private cloud over the internet. Users are relieved from the burden of having high end resources at local sites from this setup.\\
\textbf{Remote access services in $NIC$}\newline
----------------------------------------------------------- \newline
\begin{algorithmic}
       \STATE  \textit{\hl{1. Assign the remote agent}}\\
       \STATE  \textit{\hl{2. Assign the vdi service}}\\
       \STATE  \textit{\hl{3. Assign the centralized desktop services}}\\
    
\end{algorithmic}
----------------------------------------------------------- \newline

\subsubsection{Virtual Farms}

The $nIC$ is spans on the different datacenter servers. The virtual farms created upon that are for respective project domains. They all are isolated from each other by means of physical servers, vlan seperation or at application level isolation. 
The shared cloud resources are common for the servers for each domain. Individual isolated cloud resources are allocated with in the virtual farms. Each virtual farm consists of set of dynamic number of virtual instances with pre-configured software stacks. These instances are grouped according to the project requiremnts. Service login with in the farm is enabled through the techniques of nis and auto-mount. A Private storage space is part of the farm is created using the NFS. All these virtual farms are managed through the centralized orchestration techniques. We can see the total look of the setup in the fig 3.
\begin{figure} [htp]
 \centering
 \includegraphics[width=90mm,height=40mm]{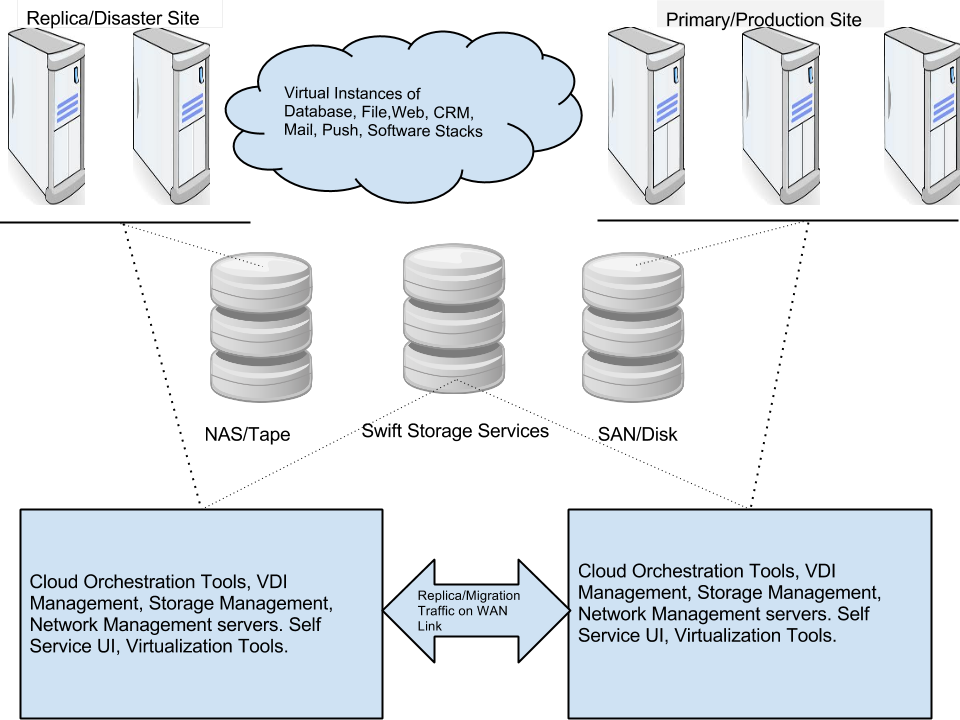}
  \caption{Virtual Farm Deployment}
\end{figure}

The primary site is totally replicated to the sceondary site and live synchronization of data is done automatically between these two sites. The second site acts as a disaster recovery site and automatically switches over to services when primary site atops working. These two sites are putten geographically different places to ensure higher reliability.

\section{Performance of the Deployment}
In this section we are presenting the deployment environment in the datacentre for various projects. The framework is deployed using the technologies mentioned in the technology stack. Build up of cloud environment by integrating these different components provisions rich service availability to the end users. We can see the management of components at each part of the framework and evaluation of such a system for efficiency measuring purposes. The utilization of different resources can be evaluated in $nIC$.

A Typical deployment consists of the following resource configuration.The Physical datacentre consists of the hundreds of servers and we constitute following configuration on them as in table 1.\\
\textbf{H/W and S/W Specification}\newline
----------------------------------------------------------- \newline
\begin{algorithmic}
       \STATE  \textit{\hl{Server           : Dell Blade/Rack server M905(15)/Quad-Core AMD Opteron(tm) Processor 8378}}\\
       \STATE  \textit{\hl{CPU              : Intel-Xeon}}\\
       \STATE  \textit{\hl{Hypervisior      : XCP/XEN}}\\
       \STATE  \textit{\hl{Orchestration    : CloudStack/OpenStack}}\\
       \STATE  \textit{\hl{BlockStorage     : SAN/Disk}}\\
       \STATE  \textit{\hl{ObjectStorage    : NFS/Disk}}\\
    
\end{algorithmic}
Virtual Instance configuration details are listed here. Further these are configured according to software stacks required by the $nIC$ users as in table 2.\\
----------------------------------------------------------- \newline
\textbf{Virtual Instance Specification}\newline
----------------------------------------------------------- \newline
\begin{algorithmic}
       \STATE  \textit{\hl{Operating System : Linux(Centos6))}}\\
       \STATE  \textit{\hl{CPU              : 16 cores(2GHz each)}}\\
       \STATE  \textit{\hl{Storage          : 2 TiB}}\\
       \STATE  \textit{\hl{Memory           : 128 GiB}}\\
       \STATE  \textit{\hl{Virtualization   : Para/Hardware}}\\
       \STATE  \textit{\hl{Networks         : 7 }}\\
    
\end{algorithmic}
----------------------------------------------------------- \newline
The Virtual farms are created according to the project requirements. The typical farm was created using the following specification as in table 3.\\
----------------------------------------------------------- \newline
\textbf{Virtual Farm Specification}\newline
----------------------------------------------------------- \newline
\begin{algorithmic}
       \STATE  \textit{\hl{Servers in the Hardware Pool : 16}}\\
       \STATE  \textit{\hl{Virtual Instances            : 500}}\\
       \STATE  \textit{\hl{Swift                        : 1 TiB}}\\
       \STATE  \textit{\hl{SAN                          : 5 TiB}}\\
       \STATE  \textit{\hl{ISO/Template Store           : 1}}\\
       \STATE  \textit{\hl{LDAP/NIS Cluster             : 1}}\\
       \STATE  \textit{\hl{Management Server Cluster    : 1}}\\
       \STATE  \textit{\hl{VDI Server Cluster           : 1}}\\
    
\end{algorithmic}
----------------------------------------------------------- \newline

Using these specification we setup a production level deployment in the datacentre for internal projects. We can see the performance of this deployment interms of seamless working state and response of the the deployment for different work loads. Various statistics are presented here are taken from the working sets of production deployments.  \\
\subsection{CPU Utilization} 
This section shows cpu resource utilization from the various kinds of project workloads in $nIC$. The setup of $nIC$ and allocation of initial allotment are based according to the work loads and showned in the fig 3 and fig 4.  \\
\begin{figure} [htp]
 \centering
 \includegraphics[width=90mm,height=40mm]{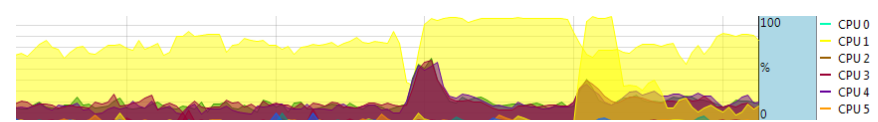}
  \caption{CPU Usage among Service Instances}
\end{figure}
\begin{figure} [htp]
 \centering
 \includegraphics[width=90mm,height=40mm]{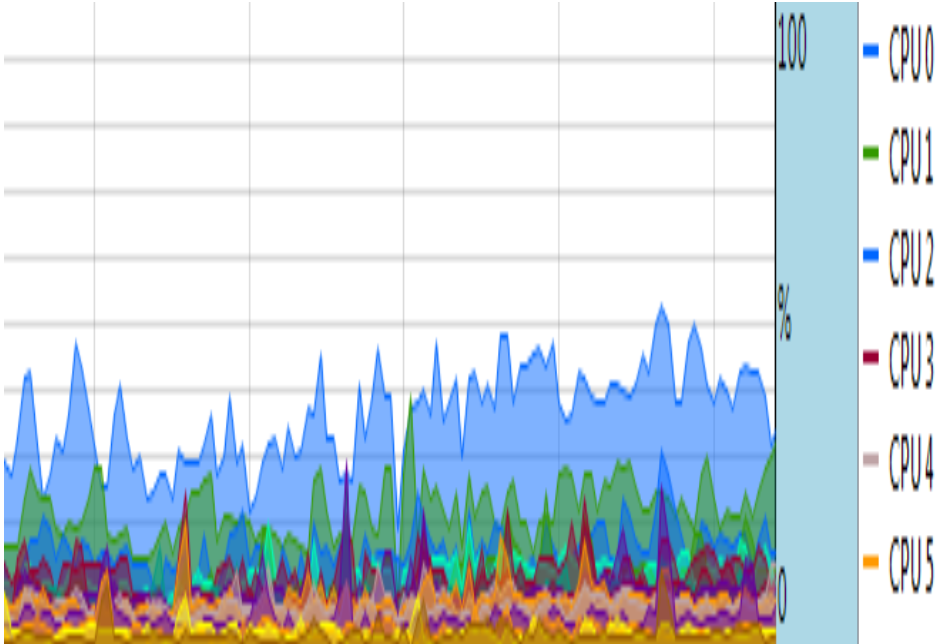}
  \caption{CPU Usage among Development Instances}
\end{figure}
The work loads of service instances which runs web, application servers  installed in the virtual instances are showned in fig 3. The template of such virtual instance is configured prior with required parameters. Mulitple vcpu(2Ghz) to service instances according to the application loads is required for optimical behaviour.\\
The work loads of development instances which runs various IDE and software stacks installed showned in fig 4 are not highly cpu intensive. The template of such virtual instance is configured prior with required parameters. The minimal allocation of cpu resource like 1 vcpu(2Ghz) for each development instance is fair choice.  



\section{Conclusion}
In this paper, we describe the design of our private cloud system $nIC$ and evaluated the typical deployment. For our system we adopted the open source software at various components of $nIC$. During our study, we have considered preformance criteria, seam-less working state and analyzed the  various project workloads in cloud. Our experimental architecture demonstrated the reference system design of internal cloud for various project workloads. We presented a production level deployment scenario and performance counters of cpu usage on this reference design. 
The difficulty of evaluating the production level deployments are lack of standardized methods for getting normalized results at each part. Further we work on disaggregation of system for better control and optimization of system behaviour. We are also in the middle of taking pragmatic results by considering efficency of system.  





\begin{thebibliography}{}

\end{thebibliography}


\begin{thebibliography}{1}

\bibitem{1}
National Informatics Centre.[Online]  Available from: $http://www.nic.in/$

\bibitem{2}
XEN and XCP(Xen Cloud Platform) [Online]  Available from:$http://www.xen.org/$
\bibitem{3}
CloudStack - An Open-Source Cloud Computing Platform.[Online]  Available from: $http://cloudstack.apache.org/$
\bibitem{4}
OpenStack  - An Open-Source Cloud Computing Platform.[Online]  Available from: $http://www.openstack.org/$
\bibitem{5}
SWIFT - An Open-Source Object Storage Platform.[Online]  Available from: $http://docs.openstack.org/developer/swift/$
\bibitem{6}
DRBD -  A Block level distributed storage system for the GNU/Linux platform. [Online]  Available from:$http://www.drbd.org/$

\bibitem{7}
GLUSTERFS -  A scale-out NAS file system. [Online]  Available from: $http://www.gluster.org/$ 


\bibitem{8}
Open vSwitch - An Software switch for cloud environments.[Online]  Available from: $http://openvswitch.org/$

\bibitem{9}
XVP - Cross-platform VNC-based and Web-based Management for Xen Cloud Platform.[Online]  Available from: $http://www.xvpsource.org/$

\bibitem{10}
VCL - A self-service system used to dynamically provision and broker remote access to a dedicated compute environment for an end-user.[Online]  Available from: $http://vcl.apache.org/$

\bibitem{10}
XenServerConfigurationLimits.[Online]  Available from:
$http://support.citrix.com/servlet/KbServlet
/download/32312-102-692726/CTX134789-XenServer6.1.0_ConfigurationLimits.pdf$


\bibitem{11}
Kumar, SMMM Kalyan, and SDMadhuKumar. "A Mobile-Cloud Paradigm for Constraint-less Computing."[Online] Available from: 
http://www.ewh.ieee.org/soc/e/sac/itee/index.php-
/meem/article/viewFile/217/224.[Accessed: 1st June 2012]

\bibitem{12}
Toor, S.; Toebbicke, R.; Resines, M.Z.; Holmgren, S., "Investigating an Open Source Cloud Storage Infrastructure for CERN-specific Data Analysis,"Networking, Architecture and Storage (NAS), 2012 IEEE 7th International Conference on, vol.,no., pp.84,88,28-30 June2012 doi:10.1109/NAS.2012.14.
\bibitem{13}
Guanghui Xu; Feng Xu; Hongxu Ma, "Deploying and researching Hadoop in virtual machines,"Automation and Logistics (ICAL), 2012 IEEE International Conference on, vol., no., pp.395,399, 15-17 Aug. 2012 doi: 10.1109/ICAL.2012.6308241. 




\end{thebibliography}
%

\end{document}